%% file: paper.tex
\documentclass[10pt,conference,letterpaper]{IEEEtran}
\IEEEoverridecommandlockouts 
\pdfoutput=1

\usepackage{url}
\usepackage{cite}
\usepackage{amssymb}

\usepackage{amsthm}
\usepackage{graphicx}
\usepackage[caption=false,font=footnotesize]{subfig}

\usepackage{amsmath}
\usepackage{xspace}

\usepackage{fixltx2e}

\usepackage[colorinlistoftodos]{todonotes}

\usepackage{mathdefs}

\newcommand{\lhs}{\emph{l.h.s.}}

\newcommand{\ie}{\emph{i.e.,}}
\newcommand{\eg}{\emph{e.g.,}}
\newcommand{\ea}{\emph{et al.}}

\newcommand{\argmax}{\operatornamewithlimits{argmax}}
\newcommand{\argmin}{\operatornamewithlimits{argmin}}

\newtheorem{lemma}{Lemma}
\newtheorem{proposition}{Proposition}

\newtheorem{remark}{Remark}

\newcommand{\lemref}[1]{Lem.~\ref{lem:#1}}
\newcommand{\prpref}[1]{Prop.~\ref{prp:#1}}

\newcommand{\remref}[1]{Rem.~\ref{rem:#1}}

\renewcommand{\eqref}[1]{(\ref{eq:#1})}
\newcommand{\secref}[1]{\S\ref{sec:#1}}
\newcommand{\figref}[1]{Fig.~\ref{fig:#1}}

\begin{document}

\title{Delay Minimizing User Association in Cellular Networks via Hierarchically Well-Separated Trees}

\author{%
  Jeffrey~Wildman, Yusuf~Osmanlioglu, Steven~Weber, and Ali~Shokoufandeh%
  \thanks{%
    J. Wildman and S. Weber are with the Electrical and Computer Engineering Department, Drexel University, Philadelphia, PA, USA (email: wildman@drexel.edu and sweber@coe.drexel.edu).%
  }%
  \thanks{%
    Y. Osmanlioglu and A. Shokoufandeh are with the Department of Computer Science, Drexel University, Philadelphia, PA, USA (email: osmanlioglu@drexel.edu and ashokouf@cs.drexel.edu).%
  }
}



\maketitle

\begin{abstract}
  We study downlink delay minimization within the context of cellular user association policies that map mobile users to base stations.
  We note the delay minimum user association problem fits within a broader class of network utility maximization and can be posed as a non-convex quadratic program.
  This non-convexity motivates a split quadratic objective function that captures the original problem's inherent tradeoff: association with a station that provides the highest signal-to-interference-plus-noise ratio (SINR) vs.\ a station that is least congested.
  We find the split-term formulation is amenable to linearization by embedding the base stations in a hierarchically well-separated tree (HST), which offers a linear approximation with constant distortion.
  We provide a numerical comparison of several problem formulations and find that with appropriate optimization parameter selection, the quadratic reformulation produces association policies with sum delays that are close to that of the original network utility maximization.
  We also comment on the more difficult problem when idle base stations (those without associated users) are deactivated.
\end{abstract}

\begin{IEEEkeywords}
  Cellular network, user association, delay minimization, quadratic program, linear approximation, hierarchically well-separated trees.
\end{IEEEkeywords}




\section{Introduction}\label{sec:introduction}

\subsection{Motivation}\label{sec:motivation}

Kleinberg and Tardos \cite{KleTar2002} investigated linearization of quadratic terms in the context of the \textit{metric labeling} problem.
Given a set of objects $P$ and a set of labels $L$ with pairwise relationships defined among the elements of both sets, metric labeling assigns a label to each object by minimizing a cost function involving both separation and assignment costs.
Separation costs penalize assigning loosely related labels to closely related objects, while assignment costs penalize labeling an object with an unrelated label.
In this paper, we seek to relate the metric labeling problem to the user association problem in a wireless cellular network, where association costs between an object (mobile user, MU) and label (base station, BS) are inversely proportional to achievable data rates between the object-label pair, while separation costs between labels can be thought of as penalizing traffic flows across backhaul links connecting BSs.
In the approximation algorithm of Kleinberg and Tardos, the label distance metric is first embedded into a \emph{hierarchically well-separated tree} (HST), which simplifies separation cost estimation.
Embedding is commonly used for tackling intractable combinatorial problems involving geometrical data~\cite{GupNewRab2004}.
Approximating the solution in such cases can be done in polynomial time once data is embedded into tree metrics.
However, such embeddings tend to introduce distortion in most cases.
Bartal \cite{Bar1996} introduced the notion of HSTs and proved the lower bound of distortion of embedding arbitrary metrics into HSTs' to be $O(\log n)$ where $n$ is the number of nodes in the source graph.
Fakcharoenphol \ea{} \cite{FakRaoKun2004} later introduced a deterministic algorithm for embedding arbitrary graphs into HSTs' with a tight bound on distortion.

\subsection{Related Work}\label{sec:related-work}

The transition from traditional cellular networks to heterogeneous networks (HetNets) has opened up many research and design questions including user association, these are gathered and detailed by Andrews \cite{And2013} and Ghosh \ea{} \cite{GhoManRat2012}.
Of interest in this paper are user association problem formulations, their complexity, and approximability.

The objective of user association is typically maximization of user rates.
The core problem often involves combinatorial optimization by mapping MUs to BSs \cite{SanWanMad2008, YeRonChe2013, SheYu2014}. 
Methods to distribute, approximate, or heuristically solve are often the key differentiating factor, while several key model assumptions also serve to differentiate approaches to the problem.
Other approaches to cell association include using stochastic geometry to characterize the outage probability of distribution of rates of a typical user in the network \cite{JoSanXia2012, SinDhiAnd2013, LinYu2013}.

Fairness of user association schemes has been addressed by several papers \cite{SanWanMad2008, SonChoDeV2009, BejHanLi2007, KimDeVYan2012}.
Sang \ea{} \cite{SanWanMad2008} propose a cross-layer, scheduling and load-aware algorithm to maximize the network's sum, weighted, \(\alpha\)-proportional fair utility.
Son \ea{} \cite{SonChoDeV2009} propose off- and on-line algorithms to compute handoff and association rules to achieve network-wide proportial rate fairness across MUs with \(\log\)-based utility.
Bejerano \ea{} \cite{BejHanLi2007} and Sun \ea{} \cite{SunHonLuo2014-ArXiv} explore association rules that promote max-min rate fairness across MUs.
Kim \ea{} \cite{KimDeVYan2012} propose a \(\alpha\)-optimal user association rule related to \(\alpha\)-proportional fairness \cite{MoWal2000} that provides a tradeoff between individual user rate maximization and load balancing across BSs.
Our work focuses on delay minimization, which is equivalent to utility maximization under \(\alpha=2\) proportional fairness.

User association policies may be either centralized or decentralized.
Many centralized problem formulations are NP-hard and require simplifying assumptions or techniques.
Corroy \ea{} \cite{CorFalMat2012} pose a relaxation of a centralized user association problem into one that is quasi-concave and provide an upper bound on the optimal value of the original sum-rate maximization.
Kim \ea{} \cite{KimDeVYan2012} propose a distributed iterative user association scheme whereby individual MUs and BSs take turns making association decisions (MUs) and advertising loads (BSs) and prove convergence to the optimal solution of the corresponding centralized optimization problem.
Shen \ea{} \cite{SheYu2014} develop a distributed pricing-based association algorithm that is based on the technique of using coordinate descent method on the dual of the original utility maximization problem.
Ye, \ea{} \cite{YeRonChe2013} propose a distributed user association algorithm based on primal-dual decomposition of an initially centralized network utility maximization problem.

\subsection{Contributions}\label{sec:contribution}

In \secref{maxrate}, we introduce our downlink cellular network model and pose a sum-rate maximization problem \eqref{maxrate-oneterm}.
In \secref{mindelay}, we pose an alternate sum-delay minimization problem \eqref{mindelay-oneterm} and show that it is quadratic but non-convex.
In \secref{linearization}, we propose a split-term delay minimization problem \eqref{mindelay-twoterm-relaxed} that captures user rate maximization with BS congestion minimization.
We show that the quadratic congestion term is amenable to linear approximation via HST embedding (\prpref{congestion}) and bound the distortion that this technique introduces into our problem formulation (\lemref{distortion-embedding}).
Thus, a linear approximation \eqref{mindelay-twoterm-relaxed-LP} is proposed in place of the split-term quadratic delay minimization formulation.
In \secref{deactivation}, we discuss permitting idle BSs, those without any associated MUs, to deactivate and reduce interference.
In \secref{results}, we provide a numerical comparison of the original combinatorial user association problem, related quadratic relaxation, split-term reformulation, and linear approximation.
Finally, \secref{conclusions} concludes our work.

\section{Downlink Rate Maximization}\label{sec:maxrate}

\input{section-maxrate}

\section{Downlink Delay Minimization}\label{sec:mindelay}

\input{section-mindelay}

\section{Linearization of Min Delay via HST Embedding}\label{sec:linearization}

\input{section-linearization}

\section{Deactivation of Idle Base Stations}\label{sec:deactivation}

\input{section-deactivation}

\section{Numerical Results}\label{sec:results}

\input{section-results}

\section{Conclusion}\label{sec:conclusions}

In this paper, we posed delay minimization via user association as a quadratic network utility maximization program.
While this quadratic representation is in general non-convex, we proposed an alternate split quadratic objective function that attempts to capture the inherent tradeoff: association with a BS that provides the highest SINR vs.\ a BS that is least congested.
We were able to extend the technique of metric embedding via HSTs to our split-term quadratic program, in which the quadratic costs of BS congestion were linearly approximated with constant distortion.
We provided a numerical comparison of several problem formulations and found that with appropriate optimization parameter selection (\(\beta\)), the split-term quadratic formulation produced sum delays that were close to that of the original network utility maximization problem.
We also commented on the more difficult problem when idle BSs (those without associated MUs) are deactivated; in this case, the SINR from a BS to a MU is additionally a function of the association map.
Future work includes examination of the proper tradeoff between instantaneous MU rate maximization and BS congestion minimization (how to select \(\beta\)), the dependence of user association problem complexity on proportional fairness parameter \(\alpha\), and further investigation of idle BS deactivation.


%



\bibliographystyle{IEEEtran}
\bibliography{IEEEabrv,refs}


\end{document}

%% file: section-maxrate.tex
Consider a set of BSs \(\Bmc\) at locations \(\{y_a | a \in \Bmc\}\) and a set of MUs \(\Umc\) at locations \(\{y_p | p \in \Umc\}\), where both sets exist within a bounded arena \(\Amc \subset \Rbb^2\).
A user \emph{association policy} is a mapping \(f : \Umc \to \Bmc\) that assigns each MU to exactly one BS.
Although modern cellular standards allow MUs to associate with multiple BSs simultaneously, \eg{} coordinated multipoint (CoMP), in this paper we ignore this generalization.

We model the instantaneous downlink rate of MU \(p\) from BS \(a\) by the Shannon rate of their point-to-point channel, treating interference from other BSs as additive white Gaussian noise:
\begin{equation}\label{eq:instratedef}
  r_{p,a} = \log(1 + \mathsf{sinr}_{p,a}),
\end{equation}
with signal to interference plus noise ratio (SINR):
\begin{equation}\label{eq:sinrdef}
  \mathsf{sinr}_{p,a} = \frac{\rho_a g(y_a,y_p)}{\sum_{b \in \Bmc \setminus a} \rho_b g(y_b,y_p) + N},
\end{equation}
determined by the BS transmission powers \(\{\rho_a | a \in \Bmc \}\), the channel attenuation function \(g(y,y')\) between locations \(y,y'\), and the noise power \(N \geq 0\).
We model channel attenuation using large scale pathloss with exponent \(\gamma \geq 2\):
\begin{equation}\label{eq:attendef}
  g(y,y') = d(y,y')^{-\gamma} = \|y - y'\|_2^{-\gamma}.
\end{equation}

An assignment \(f : \Umc \to \Bmc\) induces a partition on the set of MUs \(\Umc\), indexed by the set of BSs: \(\Umc_a(f) = \{p | f(p) = a, p \in \Umc \}, ~ a \in \Bmc \).
Denote the resulting partition cell cardinalities as \(\kappa_a(f) = |\Umc_a(f)|\) and let \(\kappa_p(f)\) denote the cardinality of the cell to which MU \(p\) is associated under mapping \(f\), \ie{} \(\kappa_p(f) = \kappa_{f(p)}(f)\).
Each BS is assumed to multiplex its resources (\ie{} time for TDMA, frequency for FDMA) fairly across all of its associated users, so that the actual downlink rate to MU \(p\) from BS \(a\) under assignment \(f\) is:
\begin{equation}\label{eq:dlratedef}
  R_p(f) = \frac{r_{p,f(p)}}{\kappa_p(f)}.
\end{equation}
This assumption of equal allotment of resources across MUs by each BS is only a convenience; future work will investigate expanding the model to incorporate potentially non-uniform allotments, as is done in \cite{YeRonChe2013}, although we note their model recovers uniform allotments as the optimal allotment (c.f.\ Prop.\ 1). 
The sum downlink rate of the network is:
\begin{equation}\label{eq:sumdlratedef}
  R(f) = \sum_{p \in \Umc} R_p(f).
\end{equation}
A reasonable objective of the downlink user association problem might be to find the association \(f\) that maximizes the sum downlink rate of the network:
\begin{equation}\label{eq:maxrate}
  \max_f R(f).
\end{equation}

\begin{remark}\label{rem:tradeoff}
  The user association problem in \eqref{maxrate} intuitively trades off the desire for a MU to associate with its strongest BS, (\(a^*(p) \in \argmax_{a \in \Bmc} \mathsf{sinr}_{p,a}\)), vs.\ the least-loaded BS, (\(a^*(p) \in \argmin_{a \in \Bmc} \kappa_a(f)\)).
\end{remark}

We can express \eqref{maxrate} directly as a combinatorial optimization problem using \(\xbf = (x_{p,a}, p \in \Umc, a \in \Bmc)\) as the vector of binary variables that together encode the assignment \(f\).
We set \(x_{p,a}=1\) iff MU \(p\) is assigned to BS \(a\).
The set of feasible assignments under single-association is denoted \(\Xmc_{\Zbb}\):
\begin{equation}
  \Xmc_{\Zbb} \!=\! \left\{\xbf: \xbf \in \{0,1\}^{|\Umc||\Bmc|}, \sum_{a\in\Bmc} x_{p,a} = 1, \forall p \in \Umc\right\}.
\end{equation}
We can write the rate of MU \(p\) as \(r_{p,f(p)}/\kappa_p(f) = \sum_{a \in \Bmc} (r_{p,a}/\kappa_a(f)) x_{p,a}\).
Next, we can represent the occupancy of BS \(a\) as \(\kappa_a(f) = \sum_{q \in \Umc} x_{q,a}\).
Substituting both expressions into \eqref{maxrate} yields the following nonlinear integer optimization:
\begin{equation}
  \max_{\xbf \in \Xmc_{\Zbb}} \sum_{p \in \Umc} \sum_{a \in \Bmc} \frac{r_{p,a}}{\sum_{q \in \Umc} x_{q,a}} x_{p,a}, \label{eq:maxrate-oneterm}
\end{equation}
where the nonlinearity stems from BS resource multiplexing.

As Ye \ea{} \cite{YeRonChe2013} state, a brute force approach to the combinatorial user association problem has a complexity of \(\Theta(|\Bmc|^{|\Umc|})\).
Following their strategy, we work towards alternative problem formulations to \eqref{maxrate-oneterm} that are easier to approximate or estimate for larger problem instances.

%% file: section-mindelay.tex
We now consider downlink delay minimization \cite{MasRob2002}, which falls under a broader class of network utility maximization problems, particularly those that employ \(\alpha\)-proportional fairness measures \cite{MoWal2000,UchKur2011}.
When \(\alpha=2\), each MU's rate is assigned a utility equal to the negation of its reciprocal: \(-1/R_{p}(f)\).
We note that log-utility (\(\alpha=1\)) combined with fractional user association constraint relaxation, turns the network utility problem into one that is convex \cite{YeRonChe2013}.
We shall find that delay minimization results in a quadratic problem upon a similar relaxation, albeit a non-convex one.

The sum downlink delay of the network and the sum downlink delay minimization problem are:
\begin{align}
  D(f) &= \sum_{p \in \Umc} \frac{1}{R_p(f)} & &\min_f ~ D(f). \label{eq:mindelay} 
\end{align}

We now provide an equivalent combinatorial representation of the delay minimization problem, analogous to \eqref{maxrate-oneterm} for the rate maximization problem:
\begin{equation}
  \min_{\xbf \in \Xmc_{\Zbb}} \quad \sum_{a \in \Bmc} \sum_{p \in \Umc} \sum_{q \in \Umc} \frac{1}{r_{p,a}} x_{p,a} x_{q,a}, \label{eq:mindelay-oneterm}
\end{equation}
noting that it is quadratic in the assignment variables \(\xbf\).

We relax the integrality constraints in \eqref{mindelay} and let \(\Xmc_{\Rbb}\) denote the new feasible set of fractional assignments:
\begin{equation}
  \Xmc_{\Rbb} \!=\! \left\{\xbf: \xbf \in [0,1]^{|\Umc||\Bmc|}, \sum_{a\in\Bmc} x_{p,a} = 1, \forall p \in \Umc\right\},
\end{equation}
yielding a quadratic optimization problem over \(\Xmc_{\Rbb}\):
\begin{equation}
  \min_{\xbf \in \Xmc_{\Rbb}} \quad \sum_{a \in \Bmc} \sum_{p \in \Umc} \sum_{q \in \Umc} \frac{1}{r_{p,a}} x_{p,a} x_{q,a}. \label{eq:mindelay-oneterm-relaxed}
\end{equation}
We find that, in general, this problem is non-convex:
\begin{proposition}[Non-Convexity of One-Term Delay Minimization]\label{prp:mindelay-oneterm-nonconvex}
  Problem \eqref{mindelay-oneterm-relaxed} is non-convex.
\end{proposition}
\begin{IEEEproof}
  Omitted for brevity.
\end{IEEEproof}
To recap, we began with the natural but difficult to solve combinatorial optimization problem \eqref{maxrate} (equivalently, \eqref{maxrate-oneterm}), then considered the modified $\alpha=2$ proportional fair combinatorial optimization problem \eqref{mindelay} (equivalently, \eqref{mindelay-oneterm}), for which integer relaxation yields the non-convex nonlinear program \eqref{mindelay-oneterm-relaxed}.
The similar agenda in \cite{YeRonChe2013} using $\alpha=1$ and integer relaxation yielded a convex program, illustrating an important difference between $\alpha = 1$ and $\alpha=2$.
In \secref{linearization}, we consider a natural variant of \eqref{mindelay-oneterm-relaxed} which we show to be a convex program.

%% file: section-linearization.tex
While we currently do not know how to directly convexify or linearize \eqref{mindelay-oneterm-relaxed}, we propose splitting the congestion term into separate terms; an assignment cost (first term) incurred by each MU and a congestion cost (second term) measured across pairs of MUs assigned to the same BS:
\begin{equation}
  \min_{\xbf \in \Xmc_{\Rbb}} ~ (1-\beta) \sum_{p \in \Umc} \sum_{a \in \Bmc} \frac{1}{r_{p,a}} x_{p,a} + \beta \sum_{a \in \Bmc} \sum_{p \in \Umc} \sum_{q \in \Umc} x_{p,a} x_{q,a} \label{eq:mindelay-twoterm-relaxed}
\end{equation}
where \(\beta \in [0,1]\) controls the relative weighting between assignment and congestion costs.
The cost of assigning MU \(p\) to a BS \(a\) is proportional to the instantaneous delay, \(1/r_{p,a}\).
Attempting to minimize assignment costs \(\beta = 0\) would result in assigning each MU to the BS providing the lowest instantaneous delay.
The cost of congestion is proportional to the number of MU-pairs associated with the same BS; \ie{} for each MU pair \((p,q)\), we pay a cost of \(\sum_{a \in \Bmc} x_{p,a}x_{q,a}\).
Assigning all MUs to one BS will make the sum \(\Theta(|\Umc|^2)\), while attempting to minimize this congestion cost would result in an even distribution of MUs across BSs.
Both cost terms maintain the tradeoff (\remref{tradeoff}) between associating with the strongest BS vs.\ associating with the least congested BS.

This formulation \eqref{mindelay-twoterm-relaxed} is in part motivated by \cite{YeRonChe2013}, where the choice of a logarithmic utility function naturally results in a similar split-term formulation.
The split term formulation provides two benefits: \emph{i)} the formulation becomes convex, and \emph{ii)} the formulation is amenable to linear approximation via HST embedding.

\begin{proposition}[Convexity of Two-Term Delay Minimization]\label{prp:mindelay-twoterm-convex}
  Problem \eqref{mindelay-twoterm-relaxed} is convex.
\end{proposition}
\begin{IEEEproof}
  Omitted for brevity.
\end{IEEEproof}

We now detail steps taken to restate \eqref{mindelay-twoterm-relaxed} in linear form \eqref{mindelay-twoterm-relaxed-LP} by reformulating the quadratic congestion term.
In its current form, each pair of MUs $p$ and $q$ assigned to BS $a$ will contribute to the overall congestion sum.
Instead, we define the congestion term using a unit distance complete graph $G=(\Bmc,E)$ whose vertices $\Bmc$ correspond to BSs.
Minimizing the congestion is then formulated as assigning MUs to BSs in $G$ such that sum of the pairwise distances among MUs as measured in $G$ is maximized.

\begin{proposition} \label{prp:congestion}
  A set of MUs $\Umc$ can be uniformly distributed over a set of BSs $\Bmc$ by solving:
  \begin{equation} \label{eq:congestion-alternative}
    \max_{\xbf \in \Xmc_{\Zbb}} \quad \frac{1}{2} \sum_{p \in \Umc} \sum_{q \in \Umc} \sum_{a \in \Bmc} \sum_{b \in \Bmc} d'(a,b) x_{p,a} x_{q,b}
  \end{equation}
  where $d'(a,b)$ is the distance metric defined over graph $G$.
\end{proposition}
\begin{IEEEproof}
  Omitted for brevity.
\end{IEEEproof}

Using an argument similar to~\cite{KleTar2002}, the quadratic term \eqref{congestion-alternative} can be linearized by using the embedding of $G$ into an HST $\Tmc$ as follows:
\begin{align}
  \max_{\xbf \in \Xmc_{\Zbb}} \quad &\frac{1}{2}\sum_{\substack{e = (p,q)\\ p,q \in \Umc}} \sum_{T\subseteq\Tmc} l_{T} \bar{x}_{eT} \label{eq:congestion-HST}\\
  \text{s.t.} \quad
  & |x_{pT} - x_{qT}| = \bar{x}_{eT},\ e=(p,q), T\subseteq\Tmc \nonumber.
\end{align}

Here $\bar{x}_{pT}$ is the probability of MU $p$ being assigned to a BS located in subtree $T$, \ie{} $x_{pT} = \sum_{a \in L(T)} x_{p,a}$, where $L(T)$ is the set of BSs included in subtree $T$.
We also note that $d'(a,b)$ in the formulation \eqref{congestion-alternative} is replaced by $\sum_{T} l_T |x_{pT} - x_{qT}|$.

We can now restate \eqref{mindelay-twoterm-relaxed} using the minimization form of \eqref{congestion-HST} as follows:
\begin{align}
  \min_{\xbf \in \Xmc_{\Zbb}} \quad & (1-\beta) \sum_{p \in \Umc}\sum_{a \in \Bmc} \frac{1}{r_{p,a}} x_{p,a} - \beta \sum_{\substack{e = (p,q)\\ p,q \in \Umc}} \sum_{T\subseteq\Tmc} l_{T}  \bar{x}_{eT} \label{eq:mindelay-twoterm-relaxed-LP} \\
  \text{s.t.} \quad
  & |x_{pT} - x_{qT}| = \bar{x}_{eT} ,\ e=(p,q), T\subseteq\Tmc \nonumber
\end{align}

Relaxing the integrality conditions to $\xbf \in \Xmc_{\Rbb}$ in \eqref{mindelay-twoterm-relaxed-LP}, we obtain a linear program.
Solving this linear program yields a fractional solution to the original problem \eqref{mindelay-twoterm-relaxed}.
The fractional solution can be rounded in the same manner as fractional assignments in the metric labeling problem \cite{KleTar2002}.
We also note that the embedding of BS graph into $\Tmc$ introduces constant distortion in the computation of congestion. 
\begin{lemma}\label{lem:distortion-embedding}
  The solution of linear program \eqref{mindelay-twoterm-relaxed-LP} has $O(1)$ distortion.
\end{lemma}

\begin{IEEEproof}
  Omitted for brevity.
\end{IEEEproof}

In summary, the split-term delay minimization formulation \eqref{mindelay-twoterm-relaxed} is convex (\prpref{mindelay-twoterm-convex}).
We also note that the linear approximation of \eqref{mindelay-twoterm-relaxed} via HST embedding may be performed with a constant distortion guarantee (\lemref{distortion-embedding}).

%% file: section-deactivation.tex
In the problem formulations discussed thus far, the downlink interference seen by MU \(p\) when associated with BS \(a\) is independent of the assignment \(f\).
That is, although MU \(p\)'s association via \(f\) determines which BS carries its signal, the interference from the other BSs is assumed fixed.
Although this independence is reasonable when all BSs have a non-empty associated set of MUs (\(\kappa_a(f) > 0, \forall a \in \Bmc\)), it is unreasonable for a BS to be assumed to transmit energy as interference when its association set is empty.
The occurrence of empty association sets may be highly likely in heterogeneous networks containing a large number of small femtocells.
If we define \(\Bmc(f) = \{a : \kappa_a(f) > 0\} \subseteq \Bmc\) as the set of active BSs under \(f\), then the SINR \eqref{sinrdef} will depend upon \(f\) as:
\begin{equation}
  \widetilde{\mathsf{sinr}}_{p}(f) = \frac{\rho_{f(p)} g(y_{f(p)},y_p)}{\sum_{b \in \Bmc(f) \setminus f(p)} \rho_b g(y_b,y_p) + N},
\end{equation}
where the interference in the denominator now only comes from active BSs \(\Bmc(f)\).

The instantaneous rate \eqref{instratedef} becomes \(\tilde{r}_{p}(f) = \log(1 + \widetilde{\mathsf{sinr}}_{p}(f))\), the downlink rate \eqref{dlratedef} to \(p\) becomes \(\tilde{R}_p(f) = \tilde{r}_p(f)/\kappa_{f(p)}(f)\), the sum downlink rate of the network \eqref{sumdlratedef} becomes \(\tilde{R}(f) = \sum_{p \in \Umc} \tilde{R}_p(f)\), and the sum downlink delay of the network \eqref{mindelay} becomes \(\tilde{D}(f) = \sum_{p \in \Umc} 1/\tilde{R}_p(f)\).
Finally, the downlink user association problems \eqref{maxrate} and \eqref{mindelay} when idle BSs don't transmit are:
\begin{align}
  & \max_f ~ \tilde{R}(f),  &  & \min_f ~ \tilde{D}(f). \label{eq:maxrate-and-mindelay-deactivate}
\end{align}

\begin{remark}\label{rem:tradeoff2}
  When idle BSs don't transmit, the MU decision-making tradeoff (\remref{tradeoff}) gains another dimension, \ie{} MUs also seek to avoid the `interference activation cost' by minimizing the number of active BSs.
  To be precise, MUs on one hand wish to load-balance themselves across BSs (\ie{} selecting associations \(f\) that induce equitable partitions \(\Umc(f)\)), while on the other hand they wish to aggregate so as to minimize interference (\ie{} selecting associations \(f\) with maximally imbalanced partitions \(\Umc(f)\)).
\end{remark}

We introduce binary BS variables \(\zbf\) that take value one iff one or more MUs are associated with the corresponding BS: \(z_a(\xbf) = \mathbf{1}\{\sum_{p\in\Umc} x_{p,a} > 0\}, \forall a\in\Bmc\).
The instantaneous rate from a BS \(a\) to MU \(p\), now a function of the association variables, can be written as:
\begin{equation}\label{eq:rate-deactivate}
  r_{p,a}(\xbf) = \log\left(1 + \frac{\rho_a g(y_a,y_p)}{\sum_{b \in \Bmc(f) \setminus a} \rho_b g(y_b,y_p)z_b(\xbf) + N}\right).
\end{equation}
Both problems in \eqref{maxrate-and-mindelay-deactivate} can be written as \eqref{maxrate-oneterm} and \eqref{mindelay-oneterm} with \(r_{p,a}\) replaced with \(r_{p,a}(\xbf)\) as shown in \eqref{rate-deactivate}.

Under the addition of variables \(z_a(\xbf)\), problem \eqref{mindelay-oneterm-relaxed} is no longer quadratic and provides stronger motivation for reformulating the problem in a more digestible form.
One might include a term that penalizes the activation of BSs that are close to one another (interference avoidance):
\begin{equation}
  \sum_{a\in\Bmc} \sum_{b\in\Bmc} f(\rho_a,\rho_b) d(a,b) z_a(\xbf) z_b(\xbf),
\end{equation}
where \(d(a,b)\) is the distance between BSs \(a\) and \(b\) and \(f(\rho_a,\rho_b)\) is a function of the transmit powers of both BSs that does not depend on \(\zbf\) or \(\xbf\).
Note that this term is quadratic in variables \(\zbf\), but nonlinear in \(\xbf\).
We leave the maturation of this formulation for future work.

In general, the deactivation of BSs improves network rates and delays due to a decrease in interference.
In the following small network example, we show that there exist scenarios in which the optimal max sum rate and min sum delay association policies do indeed benefit from BS deactivation.

\subsection{Example: Linear Network with Deactivation}\label{sec:example}

\begin{figure}[t!]
  \begin{subfloat}%
    \centering\includegraphics[width=0.475\columnwidth]{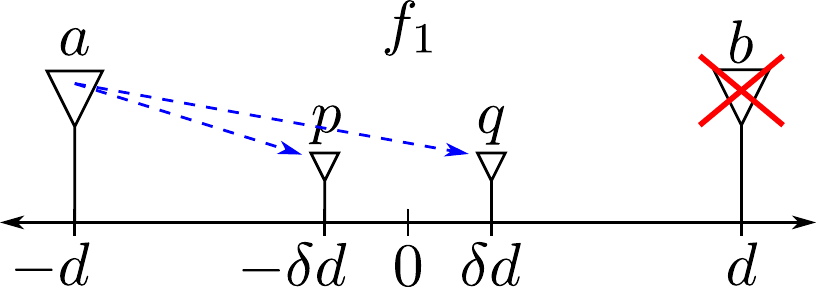}%
  \end{subfloat}%
  \hfill%
  \begin{subfloat}%
    \centering\includegraphics[width=0.475\columnwidth]{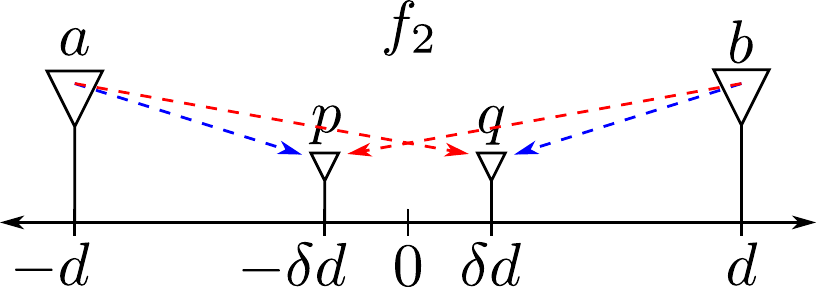}%
  \end{subfloat}%
  \caption{%
    Linear network with deactivation: assignment \(f_1:\{(p,a),(q,a)\}\) on the left, assignment \(f_2:\{(p,a),(q,b)\}\) on the right.
    Desired signals in blue, interference signals in red.
  }
  \label{fig:linearnetwork}
\end{figure}

Consider a simple network consisting of two BSs \(\Bmc = \{a,b\}\) and two MUs \(\Umc = \{p,q\}\).
Let the BSs be located at \(y_a = -d\) and \(y_b = d\) with unit power \(\rho_a = \rho_b = 1\).
Let the MUs be located at \(y_p = -\delta d\) and \(y_q = \delta d\) with \(\delta \in (0,1)\).
The parameter \(\delta\) controls how far the MUs lie away from the origin.
\figref{linearnetwork} provides the network layout for two of the possible four assignments in this example.
Observe that assignment \(f_3:\{(p,b),(q,a)\}\) is inferior to \(f_2\): both assign one MU to each BS, but \(f_2\) universally minimizes MU to associated BS distances, thus increasing rate and decreasing delay for all MUs.
Additionally, assignment \(f_4:\{(p,b),(q,b)\}\) is identical to \(f_1\) due to symmetry of the example.
Thus, it suffices to optimize problems \eqref{maxrate-and-mindelay-deactivate} under reduced assignment set \(\{f_1,f_2\}\).

\newcommand{\termone}{\left(\!1\!+\!\frac{(1-\delta)^{-\gamma}}{Nd^\gamma}\!\right)}
\newcommand{\termtwo}{\left(\!1\!+\!\frac{(1+\delta)^{-\gamma}}{Nd^\gamma}\!\right)}
\newcommand{\termthree}{\left(\!1\!+\!\frac{(1-\delta)^{-\gamma}}{(1+\delta)^{-\gamma} + Nd^\gamma}\!\right)}

\begin{proposition}[Rate Maximization on Linear Network with Deactivation]\label{prp:linear-maxrate}
  The rate maximization problem in \eqref{maxrate-and-mindelay-deactivate} on the linear network in \figref{linearnetwork} can be expressed as a threshold comparison between assignments \(f_1\) and \(f_2\):
  \begin{equation}\label{eq:maxrate-decision}
    \frac{ \termone{}^{1/2} \termtwo{}^{1/2} }{ \termthree{}^2 } \gtrless_{f_2}^{f_1} 1.
  \end{equation}
\end{proposition}

\begin{IEEEproof}
  Omitted for brevity.
\end{IEEEproof}

\begin{proposition}[Delay Minimization on Linear Network with Deactivation]\label{prp:linear-mindelay}
  The delay minimization problem in \eqref{maxrate-and-mindelay-deactivate} on the linear network in \figref{linearnetwork} can be expressed as a threshold comparison between assignments \(f_1\) and \(f_2\):
  \begin{equation}\label{eq:mindelay-decision}
    \frac{\log\termone \log\termtwo}{\log\!\termthree \! \log\!\left(\!\termone\!\!\termtwo\!\right)} \!\! \gtrless_{f_2}^{f_1} \!\! 1.
  \end{equation}
\end{proposition}

\begin{IEEEproof}
  Omitted for brevity.
\end{IEEEproof}

\begin{figure}[t!]
  \begin{subfloat}%
    \centering\includegraphics[width=0.5\columnwidth]{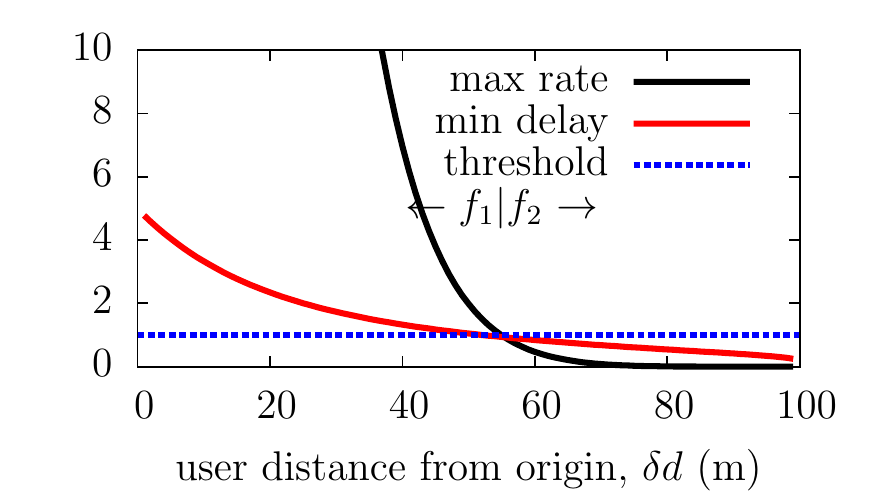}%
  \end{subfloat}%
  \hfill%
  \begin{subfloat}%
    \centering\includegraphics[width=0.5\columnwidth]{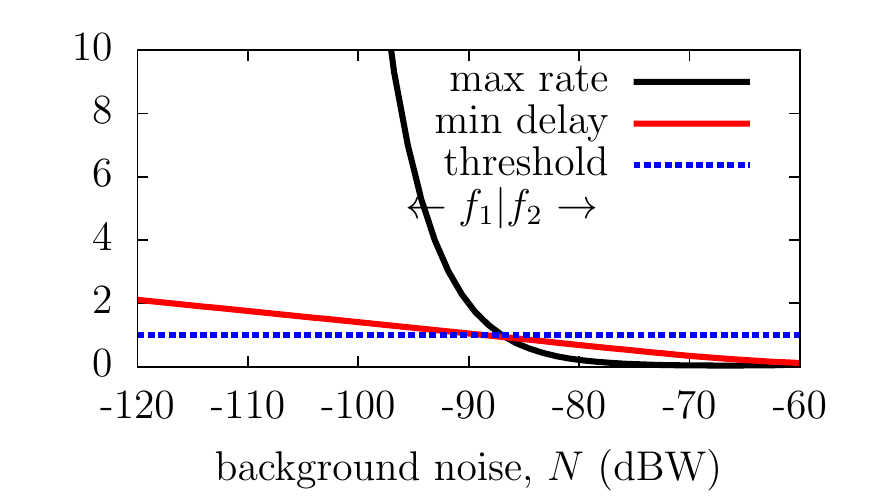}%
  \end{subfloat}%
  \begin{center}%
  \begin{subfloat}%
    \centering\includegraphics[width=0.8\columnwidth]{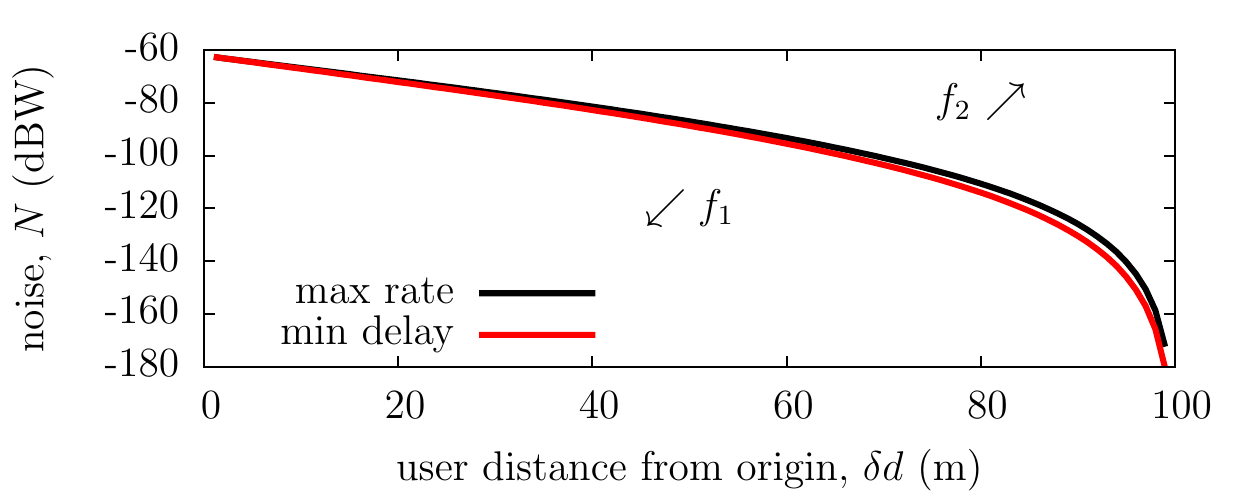}%
  \end{subfloat}%
  \end{center}%
  \caption{%
    Linear network deactivation example for \(\gamma = 3\) and \(d = 100\) meters.
    Plotted are (top-left) \lhs{} of \eqref{maxrate-decision} vs.\ \(\delta\) for constant \(N = -90\) dBW; (top-right) \lhs{} of \eqref{mindelay-decision} vs.\ \(N\) for constant \(\delta = 0.5\); and (bottom) \((\delta^*,N^*)\) pairs on the decision boundaries of \eqref{maxrate-decision} and \eqref{mindelay-decision}.
  }
  \label{fig:example-results}
\end{figure}

In \figref{example-results}, we demonstrate parameter regimes that favor either assignment \(f_1\) or assignment \(f_2\).
In the top-left plot, we see that the closer the MUs are to the origin (\(\delta\) small), assignment \(f_1\) becomes the optimal assignment.
In the top-right plot, we see that for lower background noise (\(N\) small), assignment \(f_1\) becomes the optimal assignment.
In these cases, SINR maximization/interference reduction takes priority over load balancing.
In fact, we see a partitioning of the \((\delta,N)\) parameter space based on which assignment is optimal.
Interference from additional active BSs tends to have less of an impact when \emph{i)} background noise is dominant (large \(N\)), or \emph{ii)} distances from each MU to the two BSs are highly dissimilar (large \(\delta\)).
Note the latter produces high SINRs despite the activation of both BSs.
In this regime, we see that load-balancing assignment \(f_2\) is favored over assignment \(f_1\).

%% file: section-results.tex
\newcommand{\coner}{\textsc{c1r}\xspace}
\newcommand{\coned}{\textsc{c1d}\xspace}
\newcommand{\qoned}{\textsc{q1d}\xspace}
\newcommand{\qtwod}{\textsc{q2d}\xspace}
\newcommand{\ltwod}{\textsc{l2d}\xspace}

We first provide a comparison of a subset of the problem formulations presented in this paper on a small scale network amenable to combinatorial evaluation.
We consider a \(100\) m by \(100\) m area consisting of \(4\) MUs and \(4\) BSs.
The positions of MUs and BSs are chosen to illustrate the effect of the parameter \(\beta\) on the split-term delay minimization formulation.
We consider unit BS power and a pathloss constant of \(\gamma = 3\).

\begin{figure}[t!]
  \centering%
  \begin{subfloat}%
    \centering\includegraphics[width=0.33\columnwidth]{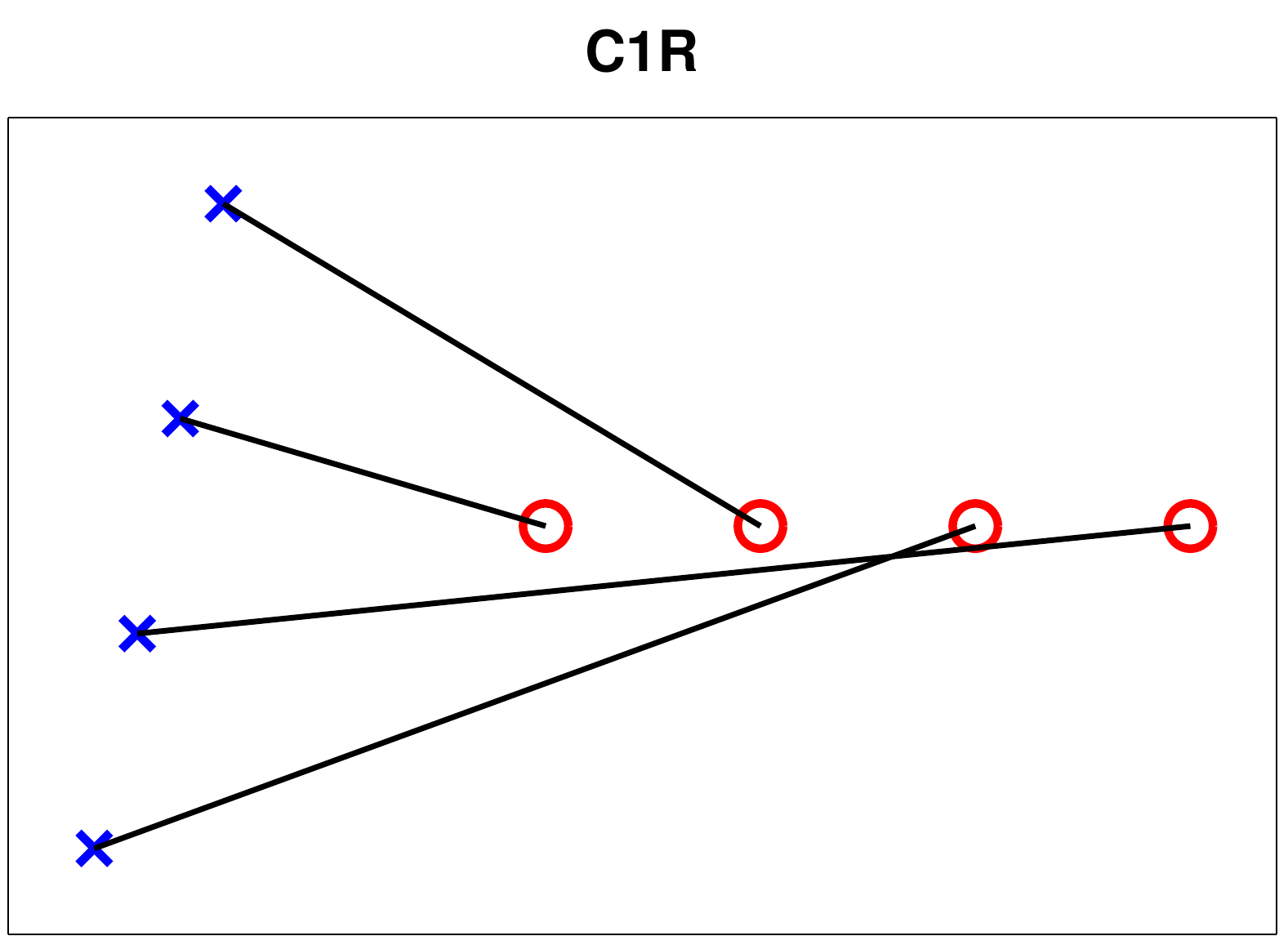}%
  \end{subfloat}%
  \hfill%
  \begin{subfloat}%
    \centering\includegraphics[width=0.33\columnwidth]{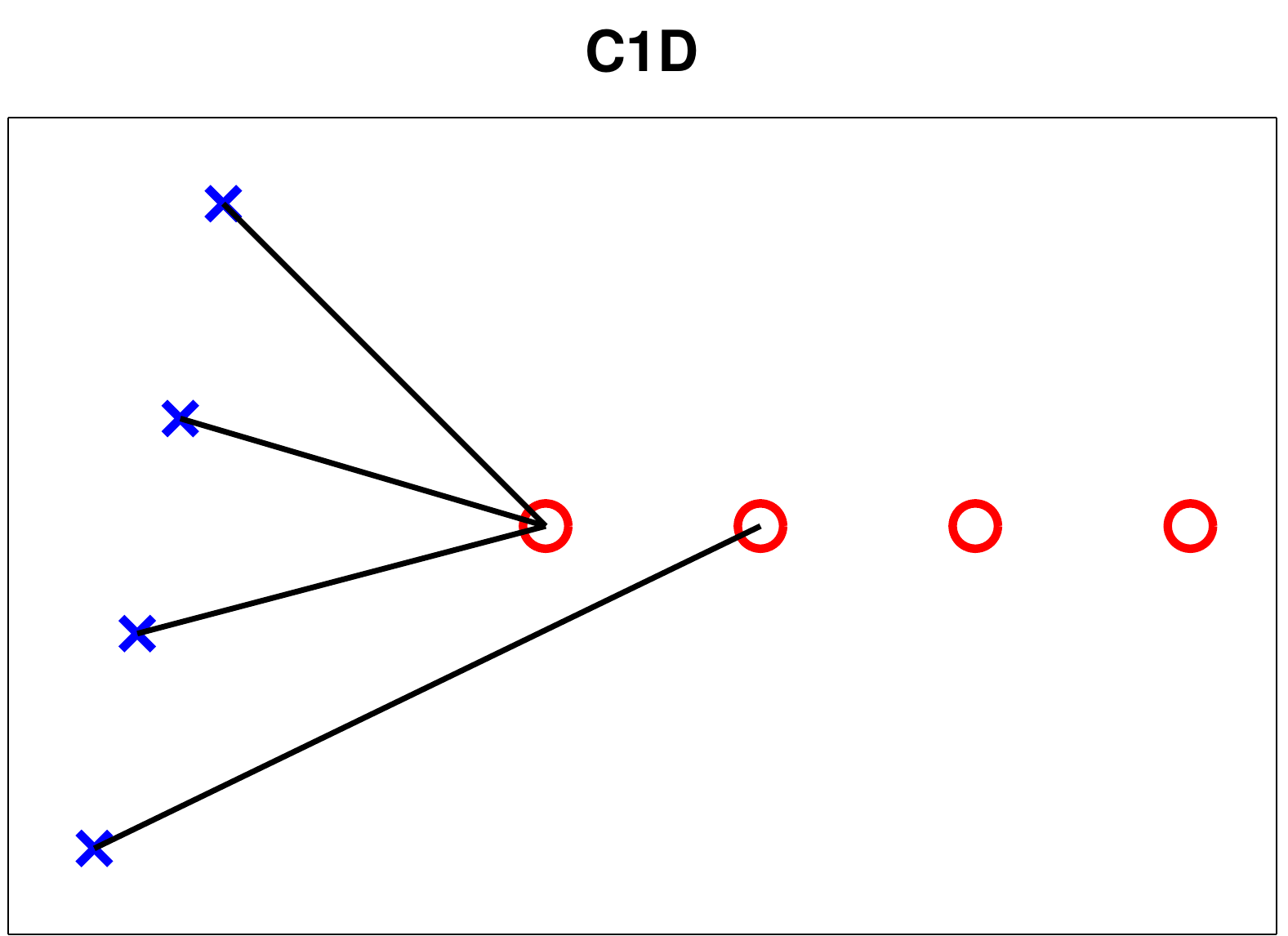}%
  \end{subfloat}%
  \hfill%
  \begin{subfloat}%
    \centering\includegraphics[width=0.33\columnwidth]{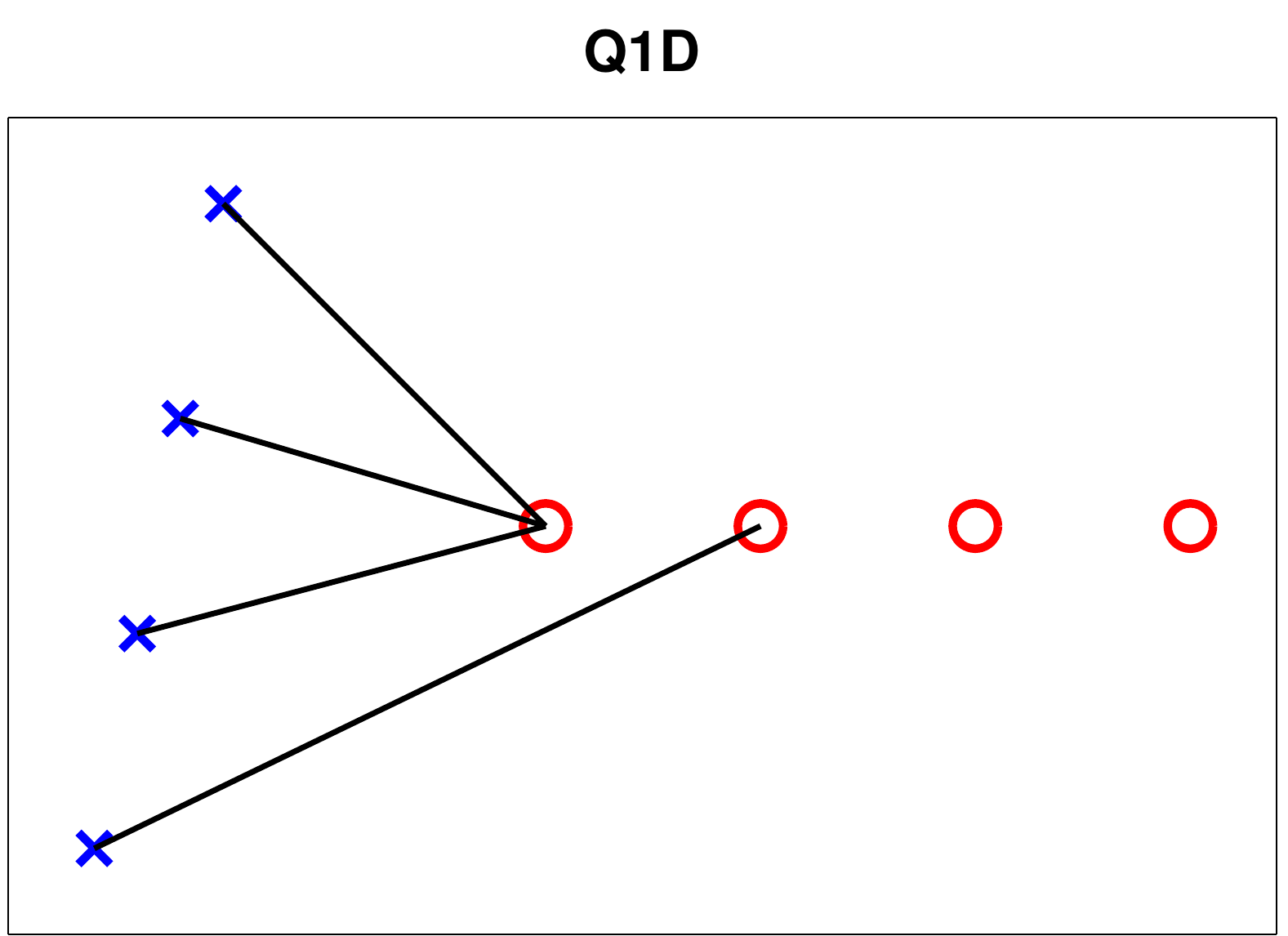}%
  \end{subfloat}\\%
  %
  \begin{subfloat}%
    \centering\includegraphics[width=0.33\columnwidth]{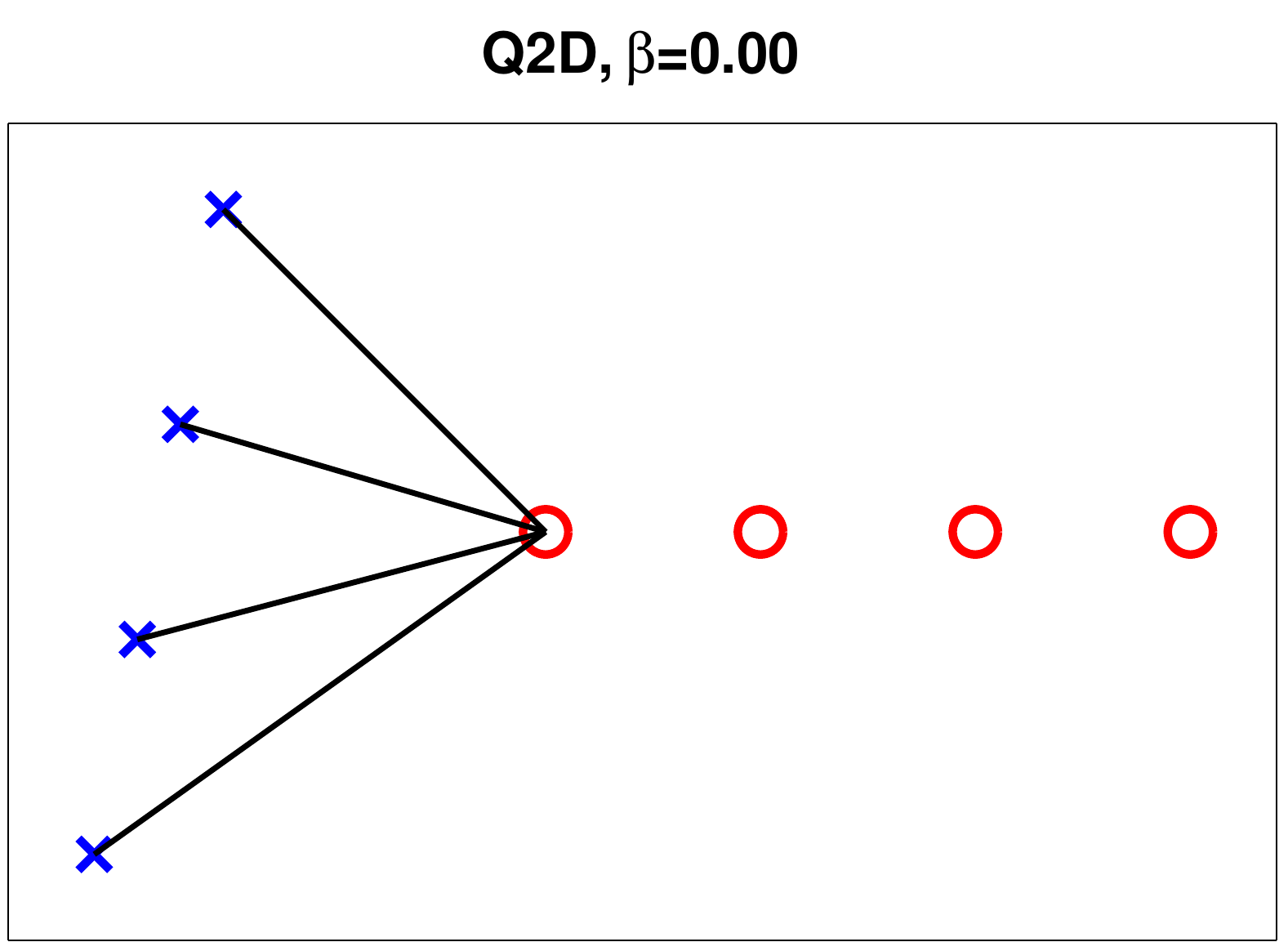}%
  \end{subfloat}%
  \hfill%
  \begin{subfloat}%
    \centering\includegraphics[width=0.33\columnwidth]{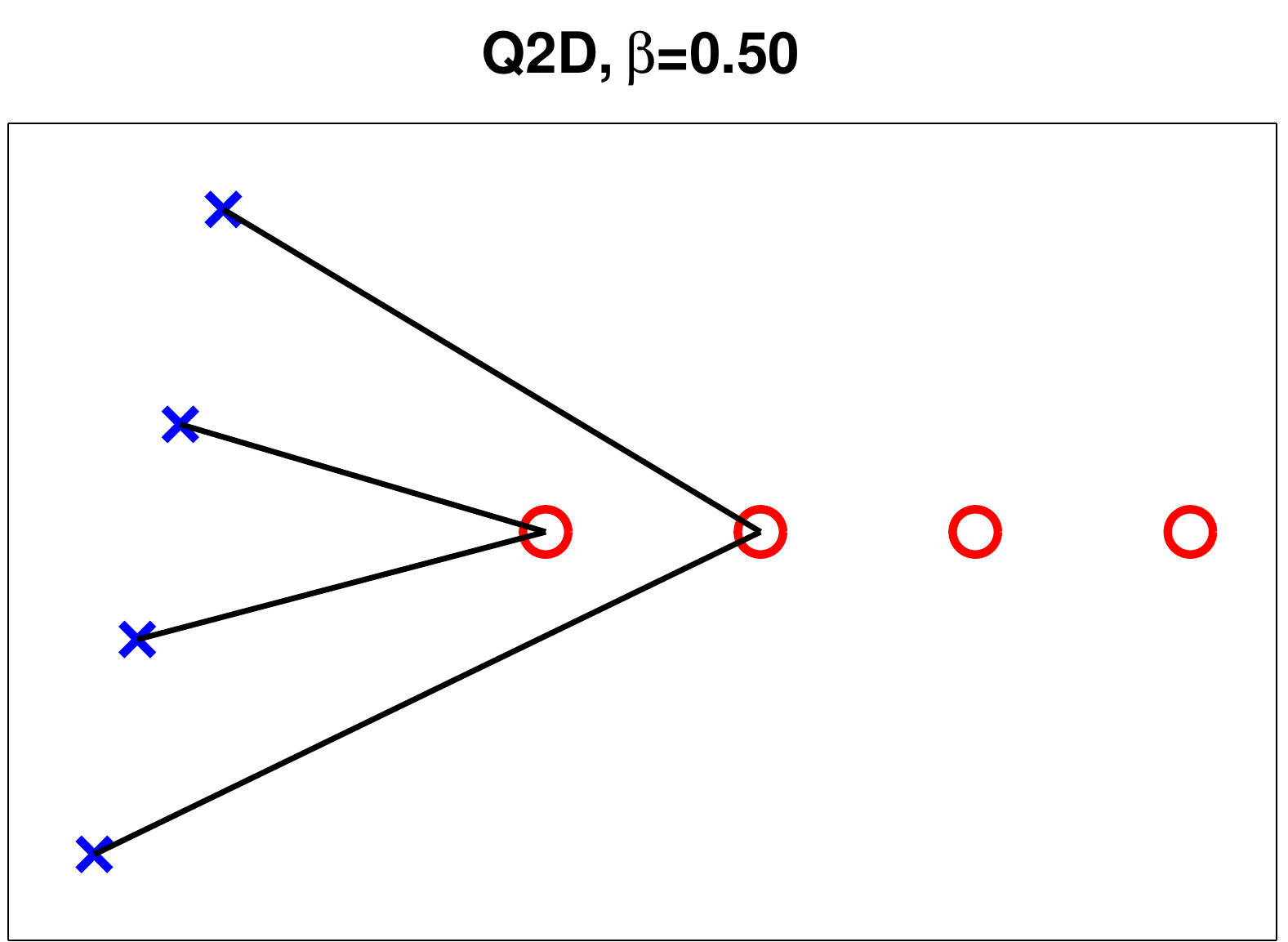}%
  \end{subfloat}%
  \hfill%
  \begin{subfloat}%
    \centering\includegraphics[width=0.33\columnwidth]{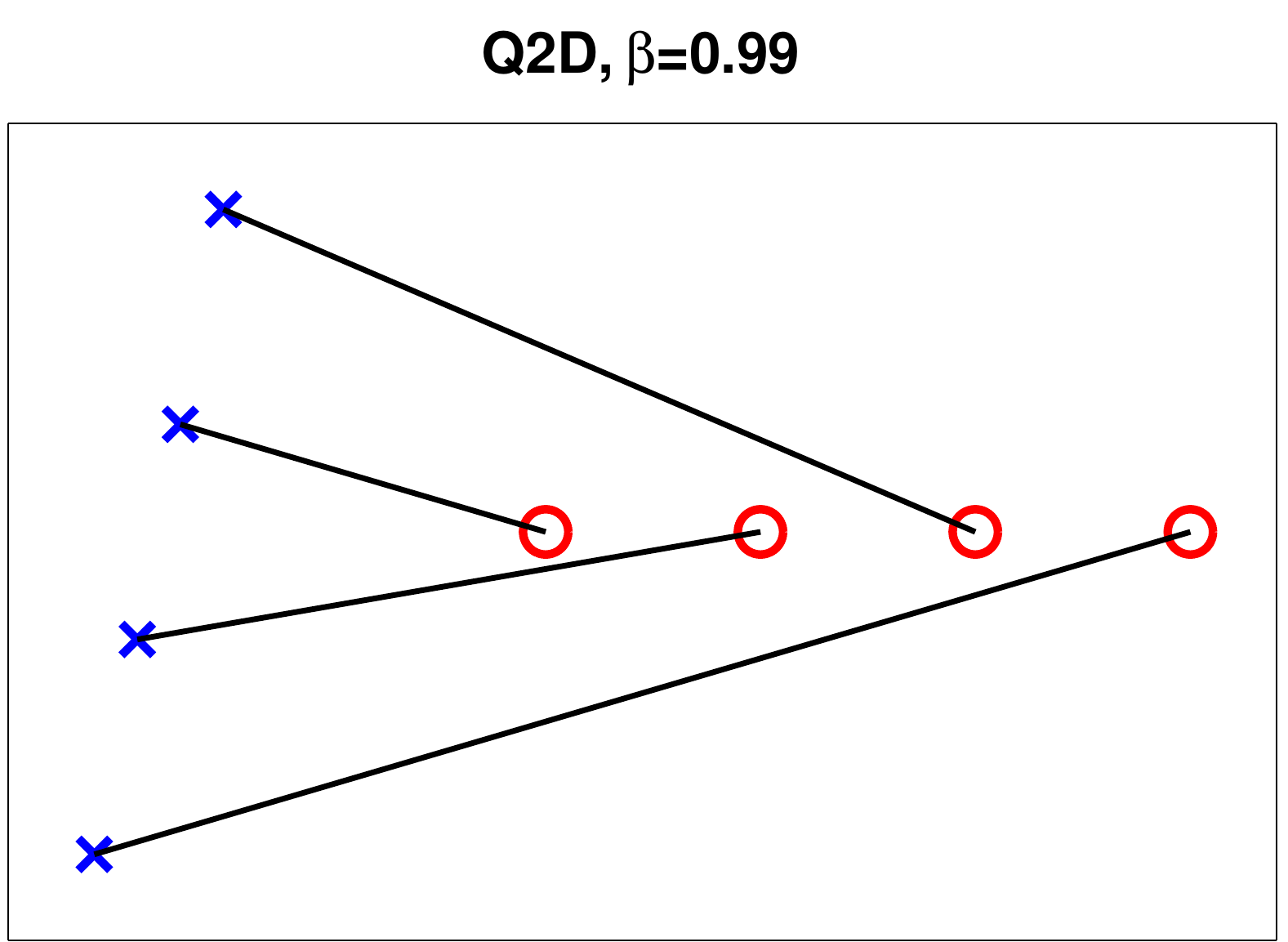}%
  \end{subfloat}\\%
  %
  \begin{subfloat}%
    \centering\includegraphics[width=0.33\columnwidth]{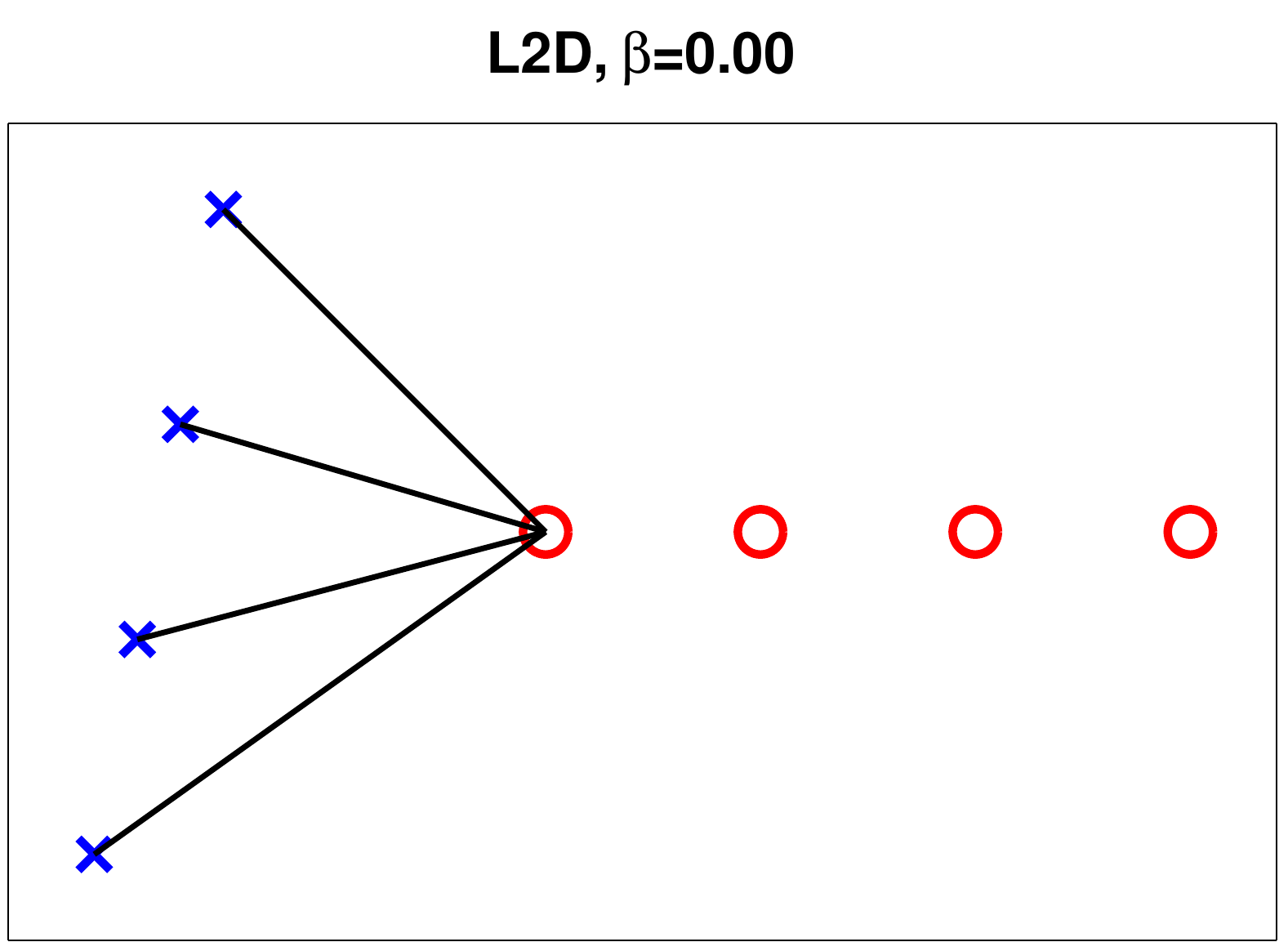}%
  \end{subfloat}%
  \hfill%
  \begin{subfloat}%
    \centering\includegraphics[width=0.33\columnwidth]{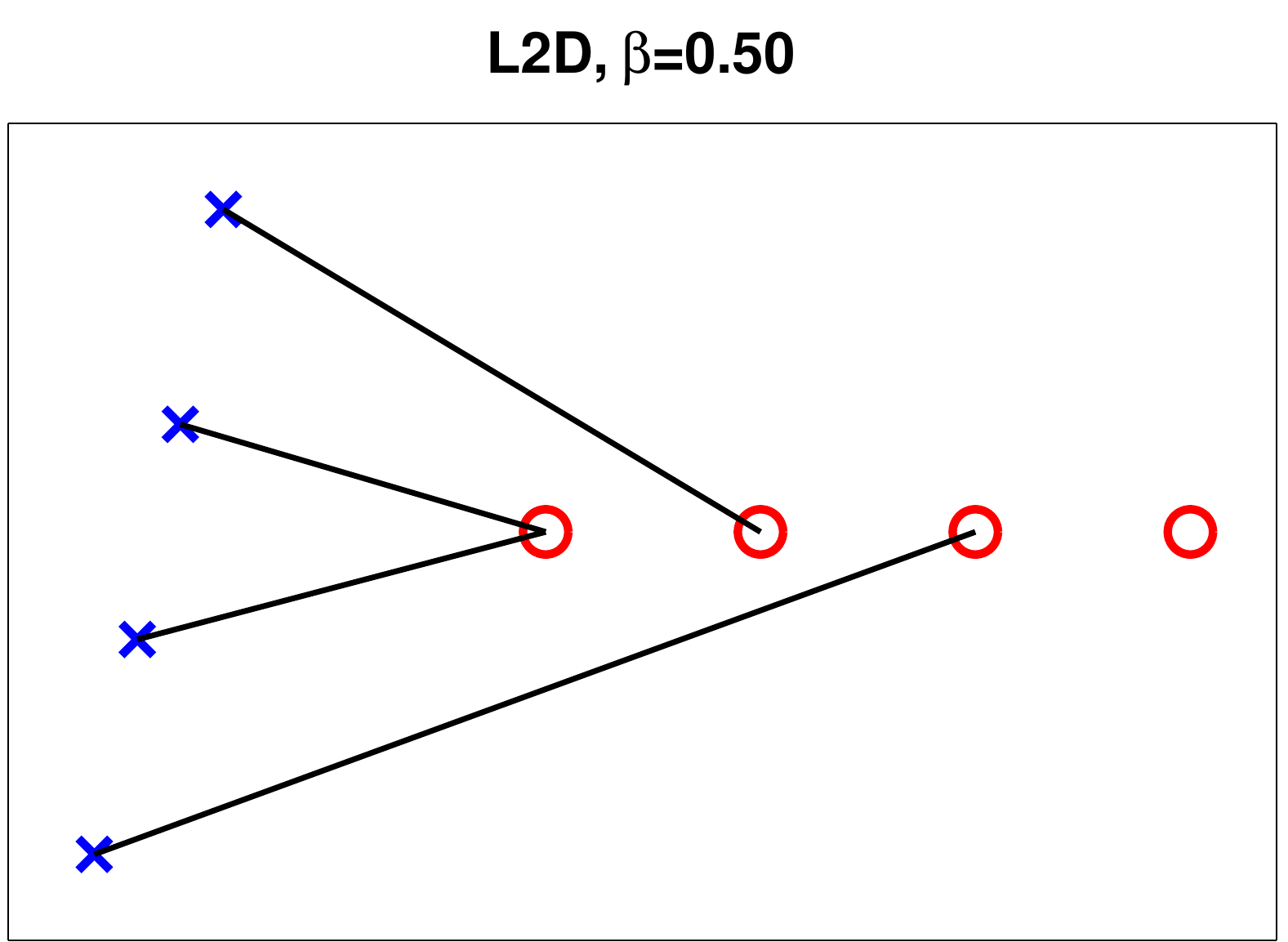}%
  \end{subfloat}%
  \hfill%
  \begin{subfloat}%
    \centering\includegraphics[width=0.33\columnwidth]{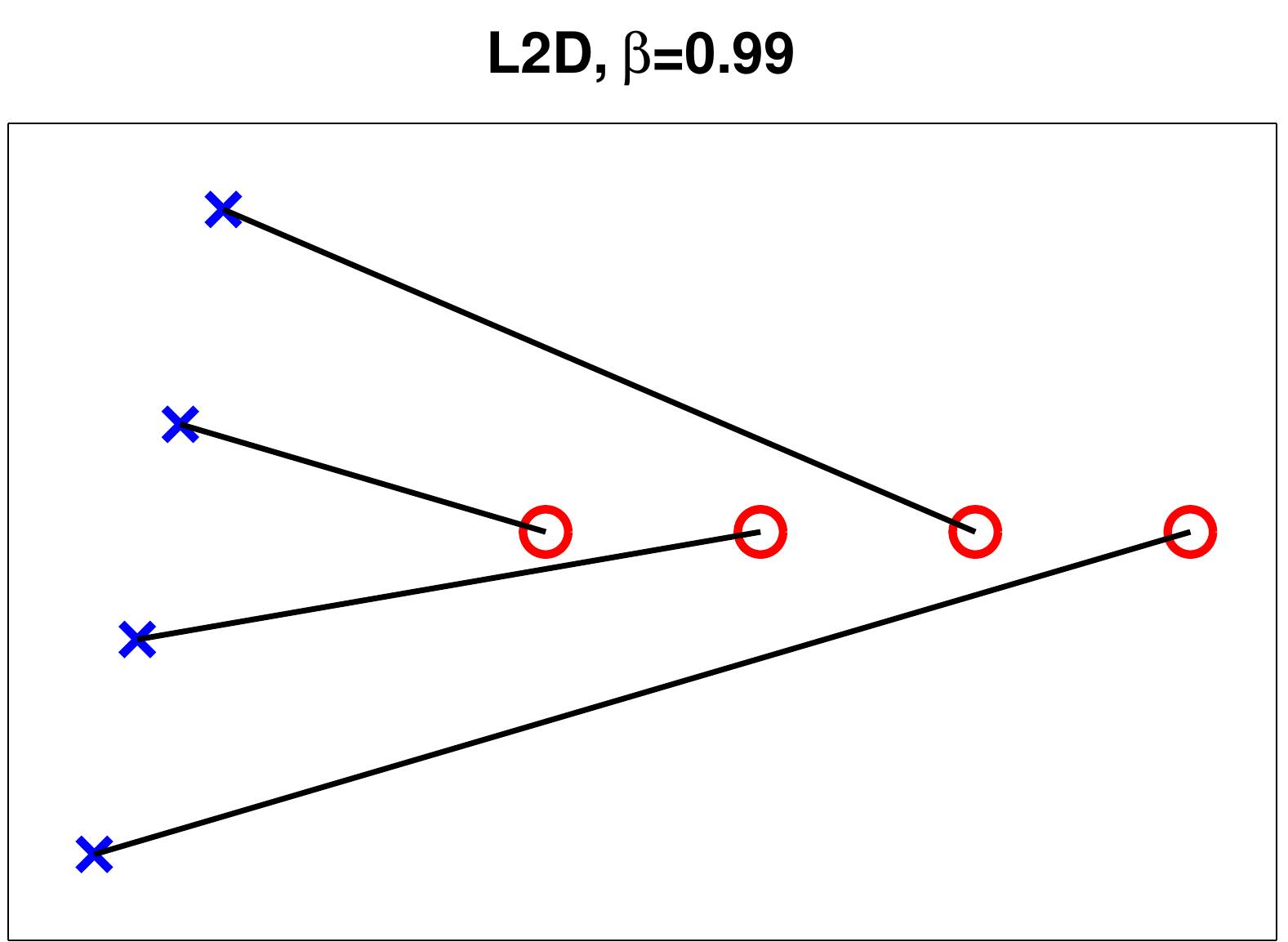}%
  \end{subfloat}%
  \caption{%
    Reported optimal user assignments, after rounding to an integer solution, on a constructed topology (base station as red circles, users as blue crosses).
    Shown are the combinatorial one-term sum rate maximization (\coner) \eqref{maxrate-oneterm} (top-left), combinatorial one-term sum delay minimization (\coned) \eqref{mindelay-oneterm} (top-middle), quadratic one-term sum delay minimization (\qoned) \eqref{mindelay-oneterm-relaxed} (top-right), quadratic two-term delay minimization (\qtwod) \eqref{mindelay-twoterm-relaxed} (middle row), and linear two-term delay minimization (\ltwod) \eqref{mindelay-twoterm-relaxed-LP} (bottom row).
    Congestion costs are controlled using \(\beta = \{0,0.5,0.99\}\) from left to right.
  }
  \label{fig:staged-optimal-assignments}
\end{figure}

For this network topology (\figref{staged-optimal-assignments}), we solve and compare several problem formulations (detailed in the caption of \figref{staged-optimal-assignments}).
Due to the non-convexity of \qoned, the reported solution may only be locally optimal.
For each of the relaxed problem formulations (\qoned, \qtwod, and \ltwod), we round the reported fractional solution into an integer solution by associating each MU with the BS for whom its fractional association was largest.

\figref{staged-optimal-assignments} displays the assignment solutions reported by each problem formulation for the constructed network topology.
In this case, \coner (top-left) avoids congestion at the BSs completely.
Alternately, \coned (top-middle) and \qoned (top-right) avoid the high-delays associated with longer MU-BS distances.
Finally, \qtwod and \ltwod (middle and bottom rows, resp.) highlight the tradeoff between instantaneous delay term minimization \(\beta = 0\) and congestion term minimization \(\beta = 0.99\).
Almost-pure congestion minimization (\(\beta = 0.99\)) causes each BS to be loaded evenly with one MU each, while pure instantaneous delay minimization (\(\beta = 0\)) results in assigning MUs to BSs purely based on SINR maximization; all MUs are assigned to the single, closest BS regardless of congestion.

\begin{figure}[t!]
  \begin{subfloat}%
    \centering\includegraphics[width=0.5\columnwidth]{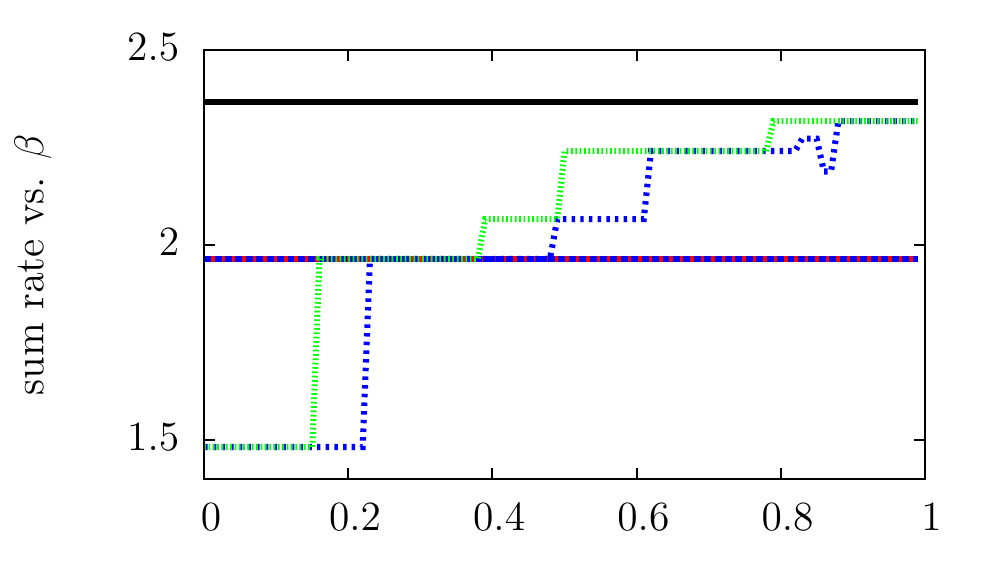}%
  \end{subfloat}%
  \hfill%
  \begin{subfloat}%
    \centering\includegraphics[width=0.5\columnwidth]{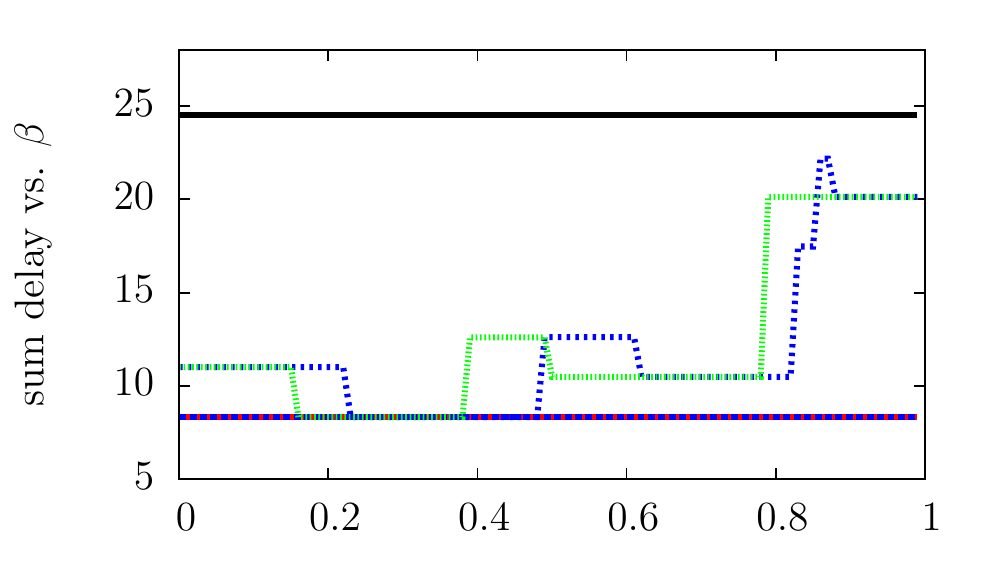}%
  \end{subfloat}\\%
  \begin{subfloat}%
    \centering\includegraphics[width=\columnwidth]{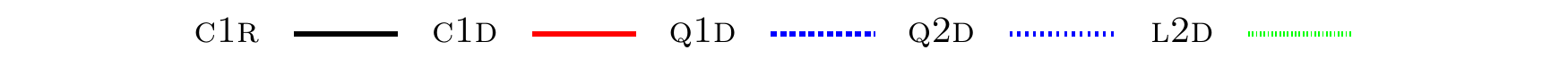}%
  \end{subfloat}%
  \caption{%
    The resulting sum rate (left) and sum delay (right) for the constructed topology shown in \figref{staged-optimal-assignments}.
    Legend corresponds to those in \figref{staged-optimal-assignments}.
  }
  \label{fig:staged-ratesdelays}
\end{figure}

An appropriate choice of \(\beta\) strikes a balance between both objectives and results in an association similar to that of \coned (top-middle).
\figref{staged-ratesdelays} compares the sum rates and delays achieved by each of the formulations shown in \figref{staged-optimal-assignments} as a function of \(\beta\).
First, note that \coner, \coned, and \qoned formulations are independent of \(\beta\).
Next, oddly, \qtwod with nearly-pure congestion minimization (\(\beta \rightarrow 1.0\)) comes close to \coner as a means of maximizing rate.
Lastly, we focus on the sum delay (right) and see that for an appropriate choice of \(\beta\), \qtwod and \ltwod can match the performance of \qoned, and all three come close to or meet the minimum sum delay reported by \coned.

\begin{figure}[t!]
  \begin{subfloat}%
    \centering\includegraphics[width=0.5\columnwidth]{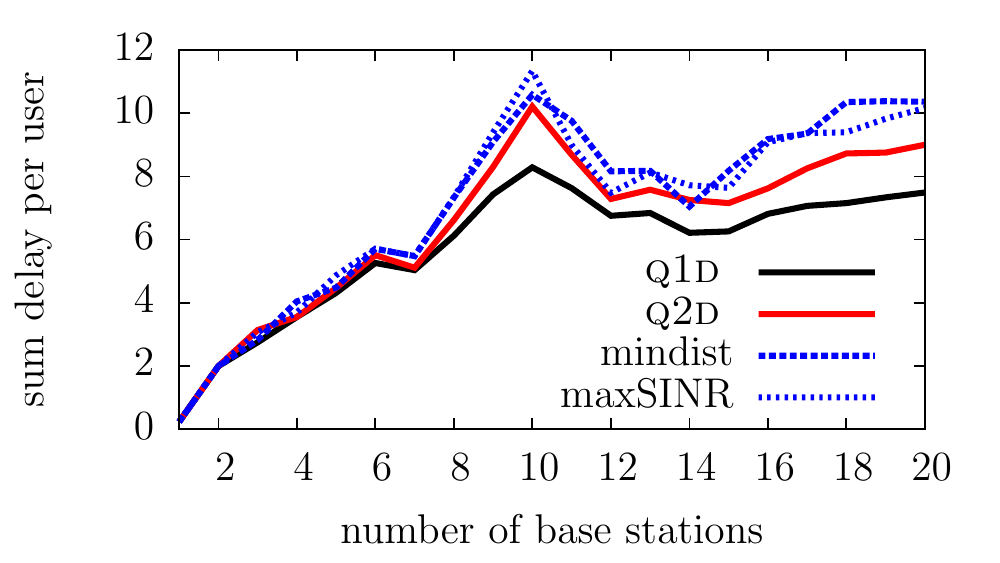}%
  \end{subfloat}%
  \hfill%
  \begin{subfloat}%
    \centering\includegraphics[width=0.5\columnwidth]{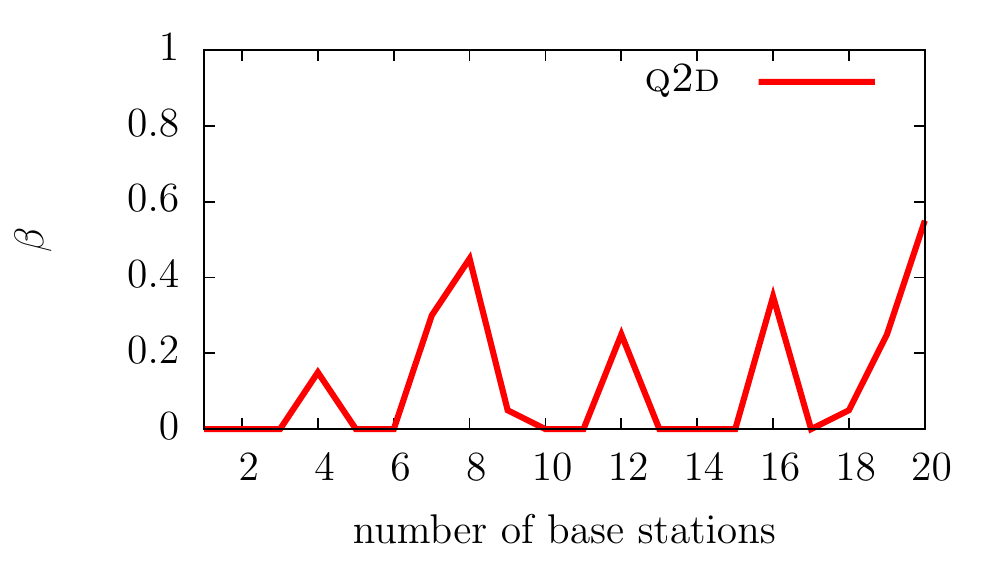}%
  \end{subfloat}%
  \caption{%
    The resulting sum delay per MU (left) as a function of the number of BSs.
    Formulation \qtwod is sampled at \(\beta\) values from \(0\) to \(1\) in increments of \(0.05\), the \(\beta\) value from this set that minimizes the sum delay (left) is reported in the figure on the (right).
    Legends correspond to those in \figref{staged-optimal-assignments} with the addition of distributed heuristic assignment policies mindist and maxSINR which minimize MU-BS distances or maximize MU-BS SINRs, respectively.
  }
  \label{fig:large-ratesdelays}
\end{figure}

Finally, \figref{large-ratesdelays} shows the delay performance of the quadratic problem formulations as the size of a randomly generated network grows from \(1\) to \(20\) BSs while the MU to BS ratio is fixed at \(5\):\(1\).
BS transmit power values are assigned uniformly at random from the discrete set \(\{50,125,250\}\).
We see that \qoned and \qtwod achieve lower sum delay (left) than commonly studied distributed heuristics \emph{mindist} and \emph{maxSINR}.
Additionally, for the sampled values of \(\beta\) for \qtwod, we found that the delay-minimizing \(\beta\) (right) tended vary between \([0,0.5]\) as the number of MUs and BSs increased.

